# The Global Landscape of Environmental AI Regulation: From the Cost of Reasoning to a Right to Green AI

KAI EBERT, European University Viadrina, Germany

BORIS GAMAZAYCHIKOV, Sustainable AI Group, France

PHILIPP HACKER, European University Viadrina, Germany

SASHA LUCCIONI, Sustainable AI Group, Canada

Artificial intelligence (AI) systems impose substantial and growing environmental costs, yet transparency about these impacts has declined even as their deployment has accelerated. This paper makes three contributions. First, we collate empirical evidence that generative Web search and reasoning models—which have proliferated in 2025—come with much higher cumulative environmental impacts than previous generations of AI approaches. Second, we map the global regulatory landscape across eleven jurisdictions and find that the manner in which environmental governance operates (predominantly at the facility-level rather than the model-level, with a focus on training rather than inference, with limited AI-specific energy disclosure requirements outside the EU) limits its applicability. Third, to address this, we propose a three-pronged policy response: mandatory model-level transparency that covers inference consumption, benchmarks, and compute locations; user rights to opt out of unnecessary generative AI integration and to select environmentally optimised models; and international coordination to prevent regulatory arbitrage. We conclude with concrete legislative proposals—including amendments to the EU AI Act, Consumer Rights Directive, and Digital Services Act—that could serve as templates for other jurisdictions.

## 1 Introduction

Despite concerns of an AI bubble, the expansion of this technology is going full steam ahead. While non-generative AI has formed part and parcel of the digital infrastructure for years, recent shifts have spurred a growing integration of various types of generative AI into ever more layers of the economic, communicative, and even artistic domains [8, 14, 69]. Reasoning models[1] are used increasingly in enterprise contexts, where they have seen a 320-fold increase in 2025 over 2024 compared to their non-reasoning counterparts [95]. In addition, more compute-intensive AI modalities, such as image and video generation models, are integrated into digital advertising at an unprecedented pace, for example in the Meta advertising suite [66]. As a case in point, take Coca-Cola: for a new advertising video for Christmas 2025, its marketing department reportedly generated 70,000 trial videos with AI [71], with no disclosure regarding the compute or energy required for these generations.

These trends result in the fact that, despite the rapid growth in adoption and deployment of AI systems, little is known about their environmental impacts. These are poised to grow significantly in coming years—both with the advent of new modalities of AI such as image and video generation, as well as the transition of existing tools, such as Web search, to using generative AI models. Despite these trends and the growing environmental costs of AI, there has not been a commensurate increase in environmental impact transparency; quite the contrary can be observed, and recent years have seen less transparency [9].

Against the background of this shifting techno-legal landscape, this paper makes three novel contributions. First, it aggregates specific numbers to document AI's rising environmental impact—this includes figures from the AI Energy Score project on the difference between reasoning and non-reasoning models that have not previously been published in an academic paper. Second, it highlights the transparency gap between rising emissions and decreasing legally mandated disclosures by a novel and comparative analysis of 10 countries and the international framework. So far, legal analysis has not focused specifically on model-level transparency and has been constrained to the EU [1, 47, 99], or the US [107, 108] in isolation. Third, based on this comparative gap analysis, we make novel policy proposals that go beyond previously published suggestions by including, for example, rights to non-generative-AI-mediated environments and to green AI models.

[1] We employ the term 'reasoning' as it is widely used, despite the potentially misleading anthropomorphism. The models do not reason in the human sense, but perform longer or different calculations in inference than standard models; see also [136].

## 2 Related Work And Case Studies

There is a growing quantity of academic scholarship that establishes theoretical approaches and frameworks for estimating the environmental impacts of AI which is relevant as grounding for our own work. We describe it in Section 2.1. We then hone in on two specific case studies that we find are representative of the environmental impacts of AI's unbridled expansion—Web search and reasoning, which we delve deeper into in Section 2.2.

### 2.1 Related Work

The first scientific article that endeavoured to quantify the energy demands and ensuing emissions of AI model training was published by Strubell et al. in 2019, establishing initial estimates and bringing this issue to the attention of the broader AI community [109]. In the years since, a growing quantity of scholarship has been dedicated to further exploring this topic, deepening and broadening its scope [10, 18, 67, 104, 131]. Notably, seminal work has identified the possible contribution of AI to climate adaptation and mitigation [17], but also to GHG emissions and toxic waste [75, 129]. The rise of new types of AI tasks and modalities also brings with it new challenges—for instance, video generation is becoming increasingly popular, while initial research has shown that it has very high energy demands even compared to previously evaluated tasks such as image generation [21].

Recent scholarship has shifted emphasis from emissions caused by training to inference [80], and taken a more holistic approach to lifecycle analysis, including the embodied emissions of manufacturing computing hardware [37, 38] as well as the relative contribution of experimentation and idle energy consumption [76, 133]. While several notable organizations such as Mistral AI [88] and Google [31] have released environmental impact disclosures on a per-prompt or per-query basis, they have refrained from providing information about cumulative usage (or even the characteristics of representative prompts), ultimately making these incomparable and difficult to interpret on a macro level [78]. However, the last years have also provided some evidence of the predicted Jevons Paradox [75], in which making models and compute cheaper and more accessible leads to increased usage, ultimately consuming more energy overall. It is difficult to prove to what extent this phenomenon is taking place, notably due to a lack of more global numbers from AI model developers and providers. What is becoming increasingly clear, however, is that the relative impact of the final model training run, the stage that had been the primary focus of initial environmental impact studies, is becoming a less relevant proportion of an AI company's total impacts—for instance, this stage only represented 5.7% of OpenAI's 2024 reported compute budget, with 'Research & Development' representing 64.5% [135].

While legal scholarship on AI's environmental effects is emerging as an increasingly prolific field of study [1, 47, 99] and attention is drawn to the legal implications of the deceptive inclusion of generative AI models [9], explicit work at the intersection of technical reporting and legal analysis remains rare [30]. Existing work has explored the intersection of legal, environmental, and ethical issues in AI [10], environmental justice-oriented algorithmic audits [103], as well as the implications of the pursuit of 'bigger-is-better' AI on society and the economy [130]. Our work builds upon this by adopting a legal lens through which we can analyse the environmental impacts of current AI growth, and chart paths for policy reform.

### 2.2 Case Studies

Given the diversity of AI models and applications, there is an increasingly wide range of environmental impacts among models that are all generally called 'AI models', from simple vector-based models that can be both trained and deployed locally without access to specialised hardware to so-called 'frontier models' that require access to thousands of cloud GPUs for training and dozens for inference. This makes it complicated to legislate AI models writ large, even for specific tasks such as Web search and text generation, which we describe in the case studies below.

### 2.2.1 Web Search

Web search has historically been one of the most commonplace applications of AI, where it has been used for several decades; it is also one of the most global AI applications in terms of roll out and scope. While previous generations of Web search used extractive approaches—i.e. ones that manipulated existing documents, converting them into vectors of numbers beforehand.[2] The current 'AI overviews' provided by search engines such as Google and Bing use generative AI approaches—i.e. producing new textual content or summaries in response to user queries, or hybrid approaches ,where information is extracted from the Web and then an LLM is used to generate the final answer that is provided to the user (called 'retrieval-augmented generation'). However, there is limited research on how these approaches compare in terms of energy given the difficulty in deploying representative systems at scale [41, 80].

It is difficult to obtain high-level estimates of the energy demands of Web search as a whole and how they have been evolving in recent years, but some numbers published by the most popular search engine provider, Google, can help get a sense of the scale of its energy demands. In a blog post from 2009, Google stated that the average (extractive) Web search used approximately 0.3 Wh of energy [51]; in a 2025 report, they report that the median text prompt for Gemini, their generative AI model, consumes 0.24 Wh, or 20% less than the number reported by Google in 2009 [31]. While it is hard to meaningfully compare these two numbers as such, we can use them to compare the cumulative impacts of Web Search in the span of 15 years—according to estimates, there were 792 billion Google searches per year in 2009 [62], and this figure is closer to 5 trillion per year in 2025 [44]. While this is not the case for all searches, if each Google Web search carried out in 2025 uses Gemini, this represents a six-fold increase in energy usage over the last 15 years.[3] Furthermore, Google reported in 2025 that their scope 1 and 2 emissions went up by 241% increase compared to 2019 (the earliest reported year), despite improvements in both hardware and software efficiency [45], illustrating the cumulative effect that increased usage can have even when the marginal energy use per query is being reduced.

### 2.2.2 Reasoning

Going a step further, AI models that carry out reasoning are increasingly being used in various text-based and agentic AI tasks. These models typically use 'chain-of-thought' (or multi-step) inference, which requires more computational steps, accompanied by a more verbose explanation of the steps involved in answering a query [cf. 136]. The AI Energy Score project, which aims to benchmark the inference energy efficiency of different types of AI models across different tasks, sheds some additional light on the impact of reasoning on text generation in general. While the February 2025 Leaderboard had a 61,848 difference between the highest and lowest energy efficiency of all of the models and tasks evaluated, this range increased to 342,822 with the December 2025 Leaderboard, which introduced reasoning models. These models were found to use, on average, 30 times more energy than models with no reasoning capabilities, significantly pushing the upper bound of the cohort as a whole[4]—some reasoning models even consumed between 150 and 700 times more energy than their base model. This is due, in part, to the fact that reasoning models tend to output longer responses, generating between 300 and 800 times more tokens than their non-reasoning equivalents given the fact that they are designed to provide textual traces of the logic they used to reply to user queries [79].

---

[2] This vectorization can be done in different ways, from sequential word-to-number mapping to more complex approaches such as GloVe [100] and word2vec [83], which reflect word co-locations and even semantics to a certain extent.

[3] Google reportedly uses backend optimizations such as caching to reduce the amount of compute they require to respond to user queries; see also [118].

[4] While the models at the maximum and minimum end of the leaderboard are not all suitable for the same task, they are all considered 'AI models' so this comparison is valid as being illustrative of the wide range within this general terminology.

In practice, reasoning models are increasingly used to respond to user queries in systems such as ChatGPT, especially for more complex queries or for premium versions of their tools [95]. While no official numbers have been reported for these models, initial estimates have found that GPT-5, the reasoning-enabled model developed in 2025 by OpenAI, uses on average 18 Wh of electricity for a medium-length prompt of 1,000 tokens [64]. Given recent estimates that ChatGPT handles 2.5 billion requests a day [105], the annual energy use of ChatGPT could equal that required for over a million average US homes.[5] However, all of these figures rely on third-party estimates and proxies. By contrast, an effective, law-based mitigation of AI's environmental impacts hinges upon access to information that would allow policymakers to create laws and regulation that reflect the realities of AI model deployment. In fact, a number of recent legal trends underscore the need for more environmental impact transparency from AI model developers and providers—we explore these in the following section.

## 3 Legal Analysis: An International Patchwork Along Three Dimensions

New modalities and types of AI models have the potential to spur a major increase in energy consumption as AI is rolled out across sectors and tools. However, the law struggles to keep pace with these developments, both at the infrastructure (e.g., data centre and energy grid) and the model level. Major jurisdictions are currently navigating the AI infrastructure race, resulting in a spectrum of policy responses that range from the EU's evolving 'hard law' reporting mandates and China's centralised green-energy directives to the United States' recent pivot toward cross-sector deregulation—and many shades in between. Overall, very few jurisdictions mandate data centre- or AI-specific energy or environmental transparency, creating a global blind spot.

Generally, the global landscape of AI environmental regulation exhibits distinct regulatory logics along three key dimensions: (1) voluntary vs. mandatory disclosures; (2) facility-level vs. model-level regulation; and (3) high vs. low environmental ambition. Concerning the first dimension, voluntary or incentive-based approaches rely on industry self-regulation, best-practice guidance, or conditional benefits such as tax credits. By contrast, mandatory disclosure regimes impose binding reporting obligations with enforcement mechanisms and penalties for non-compliance. Second, both disclosure regimes can refer to the facilities or to individual AI models. Facility-level regulation targets data centres as physical infrastructure without distinguishing AI workloads from other computational activities. Model-level regulation—though still rare—covers the environmental footprint of specific AI systems or models. These distinctions matter because facility-level metrics cannot capture the variation documented in Section 2: a data centre running energy-efficient vector search and one running reasoning-intensive inference may report identical emissions figures despite vastly different environmental impacts per query. Third, a further axis concerns regulatory ambition: some jurisdictions treat AI infrastructure as an object of environmental governance, while others subordinate environmental considerations to competitiveness objectives—or actively roll back oversight to accelerate AI deployment. The following section analyses four paradigmatic approaches found in the EU, the US, China, and Japan, with additional jurisdictions covered in the Annex.

### 3.1 The European Union: Toward Mandatory Model-Level Disclosure

The EU represents the most advanced attempt to mandate AI model-specific environmental transparency, though significant gaps remain. Current general sustainability regulations, such as the 2023 Energy Efficiency Directive (EED), the much-debated 2022 Corporate Sustainability Reporting Directive (CSRD), and the 2024 Corporate Sustainability Due Diligence Directive (CSDDD), require reporting and broader sustainability efforts. However, they lack binding targets beyond the general 2050 net-zero commitment, and the CSRD and CSDDD were significantly reduced in scope and obligations under a legal amendment that was agreed upon in the European Parliament in December 2025 [35]. Meanwhile, the EED sets up a reporting scheme under which all data centres in the EU of 500 kW or above have to report a range of sustainability indicators to a central EU database, which then publishes aggregate numbers [30]. However, in a highly controversial move, the Commission Delegated Regulation (EU) 2024/1364, after intensive industry lobbying, in its Art. 5(5) exempts consumption information of individual data centres from disclosure of public scrutiny for reasons of alleged trade secrets. This makes it all but impossible for affected communities to know the exact energy and water consumption of local data centres. This result is in clear tension with, and likely violation of, the Aarhus Convention (the UNECE Convention on Access to Information, Public Participation in Decision-making and Access to Justice in Environmental Matters), to which the EU is a signatory (see Regulation (EC) No 1367/2006). The Convention mandates public availability of environmental information, particularly also concerning emissions even in cases of an agitated secrets (Art. 4(1) and (4)(f) Aarhus Convention).

[5] According to the Energy Information Administration, the average U.S. household consumes about 10,500 kWh of electricity per year.

Concerning AI-specific regulation, the 2024 AI Act, despite declaring sustainability a main objective, remains partial, inconsistent and mostly limited to transparency provisions. Take, for instance, the energy consumption of AI training: for 'general-purpose AI models', the Act requires information on training compute and energy consumption whereas for 'high-risk AI systems' (HRAI) it only relates to training compute (Art. 53, Annex XI Sect. 1, Para. 2(d-e) vs. Art. 11, Annex IV 2(c))—with the difference and overlap between the two categories remaining unclear, and inference not mentioned at all. The 2025 General-Purpose AI Code of Practice (CoP, the Code)—which most major AI companies have signed—crucially adds requirements to document the amount inference compute, however with a big caveat, since adherence to the Code's specific requirements is voluntary. The CoP is meant to provide guidance on the AI Act's requirement to demonstrate compliance (Art. 53(4)); however, providers may at any point opt out of the Code and choose other means to demonstrate compliance—in particular if they feel the interpretation set out in the Code goes beyond the actual law. This is especially relevant as the provision on which the drafters relied to establish the requirement for inference energy merely relates to the 'process for development' (Annex XI, Sect. 1, Para. 2) [1]. It should also be noted that all of these transparency provisions apply only vis-à-vis authorities that are themselves bound by confidentiality rules (cf. Art. 21(3), 53(7), and 78(1)), while researchers, NGOs and the general public are left in the dark [1, 30].

In 2025, the EU also announced several legislative and strategic measures for early 2026 intended to rearrange its priorities on data centre regulation while also upholding sustainability efforts. This includes the EU Cloud and AI Development Act [36], which may promote green data centres through expedited permits, a Data Centre Efficiency Package [28] and a Strategic Roadmap for Digitalization and AI in the Energy Sector [29]. The regulatory logic at the EU level, hence, is one of mandatory disclosure, facility-level and partially model-level, with a mixed environmental ambition in the Omnibus package and the novel initiatives.

EU Member States have significant leeway, however, concerning various aspects of data centre regulation. One particular initiative that bears noting is a decision taken by the Irish Commission for Regulation of Utilities (CRU). In December 2025, it adopted a new large-energy-user and data-centre connection policy (Decision CRU/2025236), ending the de-facto moratorium around Dublin. As a precondition for being connected to the energy grid, any new data centre with a Maximum Import Capacity (MIC) of at least 1 Megavoltampere (MVA) must meet at least 80% of its annual electricity demand with *additional* renewable electricity generated in the Republic of Ireland, reached via a six-year 'glide path' from the date of energisation. 'Additional' here means new (or fully repowered) renewable projects not already supported under existing subsidy schemes. Compliance is typically expected via Corporate Power Purchase Agreements (CPPAs) or directly developed renewables: the CRU explicitly mentions CPPAs with new wind or solar assets or dedicated autoproducer units whose output can count towards the requirement. The same decision bundles this with resilience obligations: new data centres must have on-site (or proximate) generation or storage sufficient to cover their full demand and be able to export back to the grid in scarcity situations. Non-compliant centres face curtailment or refusal of connection. This points to an intriguing path for green data centre regulation not via legislative initiatives, such as the AI Act or the EED, but administrative decision-making which explicitly links grid connection to renewable energy targets and additionality requirements. The combination of (i) a quantified percentage, (ii) an 'additional, unsubsidised' criterion, and (iii) explicit linkage to connection rights is still distinctive to Ireland, but could serve as a model for other countries and regions, particularly given the importance of Ireland as a European home both to numerous technology corporations and data centres serving them.

### 3.2 The United States: Deregulatory Federalism

The United States exemplifies a deregulatory approach at the federal level, with environmental AI governance fragmenting across state initiatives and voluntary corporate commitments.

The framework for environmental regulation in the US rests primarily on long-standing federal legislation, such as the Clean Air Act of 1963 [124] and the Clean Water Act of 1972 [125], while more recent efforts have been hindered by political polarization, leaving introduced bills largely symbolic [121]. By contrast, individual states have been highly active in regulating AI, including with respect to the energy consumption of data centres and to transparency [70], but not at the model level. Even California's AB 2013 (effective January 1, 2026) requires developers to only post 'high-level summaries' of training data, including dataset sources, volume, and collection methods [20]. While this mandates transparency into AI development processes, the law omits energy consumption and compute from required disclosures. However, as of 2026/2027, California's Climate Corporate Data Accountability Act (SB 253) does require companies doing business in California and with total annual global revenues exceeding $1 billion to disclose their Scope 1, Scope 2, and Scope 3 emissions, which for AI companies can be argued to clearly include emissions from data centres. More generally, however, the second Trump administration's push for deregulation—relying mainly on executive action—has created political tension between the federal and state levels, culminating in a federal push for pre-empting state-level AI regulation [115].

Whereas the Biden administration sought to balance technological leadership with accountability—most notably through measures such as Executive Orders 14110 [126] and 14141 [114], and substantial investments in renewable energy and infrastructure under the bipartisan Infrastructure Investment and Jobs Act (IIJA) [90] and the Inflation Reduction Act (IRA) [123]—the second Trump administration has marked a decisive shift in priorities. It emphasises fossil energy, broad regulatory rollbacks, deregulation in the energy and environmental sectors, and reductions in agency oversight, with the facilitation of the AI race as a central objective. Agencies have been directed to scale back enforcement accordingly, with the EPA pursuing what was officially labelled the 'biggest deregulatory action in U.S. history' [122], the SEC halting enforcement of federal climate disclosure rules [127], and the FTC reducing scrutiny where it could unduly burden AI innovation [117]. This approach was reflected in a series of executive actions, including EO 14179, Removing Barriers to American Leadership in Artificial Intelligence [116], the AI Action Plan 2025 [117], and Executive Order 14318, Accelerating Federal Permitting of Data Centre Infrastructure [113]. Efforts to pre-empt conflicting state AI regulation were, after strong opposition in the legislature [74, 128], ultimately formalised in Executive Order 14365, Ensuring a National Policy Framework for Artificial Intelligence [115]. The overall regulatory logic is voluntary, facility-level only (through general environmental law), cross-sectoral, and with clearly decreasing environmental ambition.

### 3.3 China: State-Directed Green Infrastructure

The AI race falls amid diverging paths of the U.S. and China in energy strategy: while the U.S. swings fully behind fossil energy, China is making intensive investments in green energy and a nuclear power build-out—while also still relying heavily on coal. Hence, China pursues AI infrastructure expansion within a state-directed framework that emphasises low-carbon energy deployment but not transparency.

Sustainable AI infrastructure has become a recurring national priority across industrial and policy documents, from the 2015 Made in China 2025 [134] and the 2017 New Generation AI Development Plan [132] to the Special Plan for New Data Centre Layout ('Eastern Data, Western Computing') launched in 2021 to shift energy-intensive data centres toward resource-rich western Chinese regions [138]. This emphasis has been reinforced in the 14th Five-Year Plan (2021–2025) [19], the Special Action Plan for Green and Low-Carbon Development of Data Centres issued in July 2024 [112], and ongoing standard-setting efforts by the Ministry of Industry and Information Technology, including its October 2025 consultation on computing power standards that highlights green and low-carbon development [27]. Draft recommendations for the 15th Five-Year Plan (2026–2030), presented by Xi Jinping in October 2025 and formally approved in March 2026, further confirm green development as a core pillar, signalling that China increasingly frames its AI ambitions within a long-term energy transition narrative rather than treating energy and AI as separate policy domains [40]. Outside of these substantive measures, however, no particular transparency requirements for data centres or AI have been instituted. The regulatory logic involves mandatory efficiency standards, facility-level regulation, but no public transparency. The level of environmental ambition, however, is increasing.

## 3.4 Japan: Hybrid Governance

Japan combines mandatory energy reporting with voluntary strategic coordination, which produces a hybrid regime that addresses facility-level impacts but not AI-specific consumption. Under the 2022 revision of Japan's Energy Conservation Act, certain business operators that exceed defined energy-consumption thresholds—including data centres—are subject to mandatory energy reporting and transparency obligations, requiring disclosure of energy use, efficiency measures, and mid- to long-term improvement plans [63]. For data centres, this includes efforts to reduce power usage effectiveness (PUE), with a sectoral target of 1.4 by 2030, corresponding to the top 15% of performers in the 2020 reference year, backed by the possibility of fines for non-compliance [4]. However, given the long timeline and current state of data centre efficiency, this goal seems only moderately ambitious. Alongside this hard regulation, Japan has pursued a complementary, pro-innovation policy agenda linking AI growth, data centre expansion, and decarbonization. Initiatives such as the 2025 Watt-Bit Collaboration [63] and the GX Strategic Zone framework [26] emphasise coordination between infrastructure planning, renewable energy deployment, and voluntary strategic planning rather than prescriptive, AI-specific energy disclosure mandates, distinguishing Japan's approach from more demanding regulatory logics such as that of the EU.

## 3.5 Discussion

Three patterns emerge from this comparative analysis. First, facility-level regulation dominates. Most jurisdictions that address data centre energy consumption do so through economy-wide energy efficiency laws or carbon pricing mechanisms. These instruments, in some instances, capture data centres as large electricity consumers but cannot distinguish AI workloads from other computational activities, and certainly cannot differentiate between model types with vastly different per-query impacts. The Irish example, however, shows that individual countries can successfully enact rules concerning sustainability targets for data centres via administrative decision-making, in this case as a precondition for access to the energy grid.

Second, AI-specific transparency remains rare. Only the EU has enacted binding requirements for AI model energy disclosure, and these apply primarily to training rather than inference. The General-Purpose AI Code of Practice extends to inference compute, but its voluntary character limits enforceability, and data is not shared publicly. No jurisdiction requires public disclosure of inference energy consumption, compute locations, or standardised efficiency metrics that would enable cross-model comparison.

Third, regulatory objectives diverge. The EU pursues accountability through mandatory disclosure. The United States, under the second Trump administration, prioritises AI competitiveness through deregulation. China emphasises state-directed efficiency without external transparency. Japan combines binding facility requirements with voluntary AI-sector coordination. Among the other jurisdictions relegated to the Appendix, Singapore, the UAE, and Brazil approximate Japan's hybrid model by combining mandatory economy-wide obligations with voluntary sector-specific guidance, while the United Kingdom and Canada follow the US hands-off approach that prioritises facilitation over disclosure. India has not yet settled on a unified policy response to the growing energy demands of data centres and AI. These divergent approaches create opportunities for regulatory arbitrage: providers may locate compute in jurisdictions with weaker disclosure requirements while serving users globally. The absence of model-level transparency constitutes the most significant gap. As Section 2.2.2 demonstrates, reasoning models consume orders of magnitude more energy than non-reasoning alternatives for comparable tasks. Facility-level metrics—PUE targets, aggregate energy reporting, carbon intensity figures—cannot surface this variation. Users, downstream deployers, researchers, and policymakers lack the information necessary to assess AI's true environmental footprint, compare alternatives, or design effective interventions. This transparency gap motivates the policy proposals developed in Section 4.

# 4 Policy Implications

We argue that the environmental impact of AI demands a three-pronged policy approach: specific transparency mandates to address information asymmetries; user rights to rein in externalities that markets alone cannot internalise; and international coordination to mitigate regulatory arbitrage. We propose ways to operationalise these three endeavours in the sections below, contrasting the current and desired state of global policy.

## 4.1 More Concrete Measures of Transparency

To bridge the gap between the current landscape of fragmented, voluntary reporting and the necessity for rigorous environmental accountability, transparency obligations must shift from aggregate facility-level reporting to specific model-level disclosure. Currently, a dichotomy exists where hyperscalers report aggregate data that obscures AI-specific impacts, while model developers release non-standardised inference numbers that prevent meaningful comparison [77]. Effective governance requires a transition to a standardised regime where model-level and organization-level disclosures serve as the foundation for accurate global mitigation. We propose the following specific mandates to operationalise this shift.

### 4.1.1 AI Model-Level, Not Only Data Centre-Level Reporting

To operationalise this framework, transparency obligations must shift from aggregate facility-level reporting to specific model-level disclosure to enable accurate environmental impact assessment. The EU has the most far-reaching rules in this respect, but focuses on training, not inference. Hence, the EU should modify the GPAI model rules in the AI Act, specifically Art. 53(1) and Annexes XI and XII. The addition of two model-specific types of transparency—inference consumption and compute locations—is essential. While full lifecycle analysis of AI's environmental effects would require much more information [38], we limit ourselves to what remains both most urgent and realistic to implement without excessive burden on providers.

#### 4.1.1.1 Inference Model Consumption

Model providers must publicly disclose energy and water consumption during inference operations, as this phase often exceeds training in cumulative resource use [80]. This consumption constitutes an important market metric for downstream deployers who choose between different models, and for researchers and NGOs who aim to estimate AI's climate effects.

#### 4.1.1.2 Cumulative Effects of Inference Model Consumption

Beyond individual query-level disclosures, legislators should also mandate the reporting of aggregate consumption data to capture the cumulative environmental footprint of large-scale AI deployment. We propose two principal regulatory options which merit consideration.

*Option 1: Granular Operational Metrics*. Under this approach, AI system providers would be required to disclose:

(1) **Query volume**: The total number of queries processed per day (or per reporting period) for each major AI system (e.g., Gemini, Claude, ChatGPT).

(2) **Distributional energy consumption data**: Rather than relying solely on mean energy consumption per prompt, providers should report the median, mode, and standard deviations. This requirement would illuminate the substantial variation in resource consumption across different query types—from simple factual lookups to complex reasoning tasks or multimodal generation.

(3) **Cache utilization ratios**: Providers should disclose the ratio between cached responses (queries answered through retrieval of pre-computed results without new inference) and unique generations (queries that require fresh inference passes) as the former consume a fraction of the energy required for novel generations [118].

The granular approach offers significant advantages for external verification and scientific analysis. Researchers and regulators could cross-reference reported figures against independent estimates, and the detailed breakdown would enable more precise modelling of environmental impacts under different usage scenarios. However, providers may perceive these disclosure requirements to be onerous and commercially sensitive.

*Option 2: Aggregate System-Level Reporting*. Alternatively, regulators could mandate a single consolidated figure that represents the cumulative annual energy consumption, carbon emissions, and water usage attributable to each AI system or product suite. Under this model, a provider would report, for instance, that the entire Claude system consumed X terawatt-hours of electricity, generated Y metric tons of CO2-equivalent emissions, and required Z megalitres of water over the preceding calendar year. This approach offers practical advantages: it reduces compliance complexity for providers and yields easily comparable figures across competing systems. A consumer or enterprise client could directly compare the annual environmental footprint of Gemini against Claude or GPT-5 when making procurement decisions. However, the aggregate approach comes with a significant limitation. A single consolidated number is difficult to verify externally without access to the underlying operational data. Auditors and researchers would lack the granularity necessary to assess whether reported figures are plausible or to identify anomalies.

**Recommended Approach**. A hybrid mandate would combine the comparative utility of aggregate figures with the verifiability of granular data. Legislation should require AI system providers to publish both: (a) cumulative annual consumption figures for energy, carbon, and water at the system level; and (b) supporting operational metrics that include query volumes, distributional energy data, and cache utilization rates. In this way, the public and regulators would gain access to a single headline figure as well as the underlying metrics supporting auditability.

#### 4.1.1.3 Compute Locations

Geographic disclosure of compute locations enables stakeholders to assess both data sovereignty concerns and the carbon intensity of AI operations. Providers should disclose, at minimum, the compute region (e.g., us-east-1) where compute occurs for user queries that originate in each geographical region, ideally broken down by country or state, as well as if the local energy grid is utilised or if a custom ('behind the meter') energy source is used instead. This disclosure allows researchers, NGOs, downstream deployers and end users to measure or estimate where potentially sensitive data undergoes processing and what type of energy mix (renewables versus fossil fuels) typically powers a certain model in a certain region. This information can both fuel policy debates and drive meaningful user behaviour change.

#### 4.1.1.4 Visual Rating Systems

Clear visual indicators must synthesise complex environmental data into actionable information for users and deployers. Both dimensions—inference consumption and compute location—should inform AI model ratings in a clear, conspicuous, and visual manner. Such an approach provides orientation to downstream deployers and end users. The AI Energy Score framework provides a promising foundation for such ratings. It evaluates models across multiple tasks and modalities using standardised datasets and consistent hardware configurations. Its star-based system (1 to 5 stars) is inspired by energy scores for appliances such as refrigerators, translating complex energy consumption data into immediately comprehensible signals for downstream deployers and end users. Empirical studies demonstrate that such energy ratings can have a beneficial effect on informing consumers concerning the energy consumption of goods they buy or use, and spur the purchase of more energy-efficient appliances [2, 39].

#### 4.1.1.5 Towards Mandatory Disclosure Of Energy Scores And Monetary Information

Regulators should, therefore, mandate the adoption of such standardised energy scoring frameworks for AI models and systems in a layered framework to ensure comparability across providers. First, the EU AI Office, in coordination with standardization bodies such as CEN/CENELEC, should consider endorsing or building upon the AI Energy Score methodology when developing technical standards under the AI Act's GPAI provisions. Second, for consumer-facing disclosures, the AI Energy Score or a comparable visual grading system should be mandated to be displayed on the landing page of specific model providers (e.g., openai.com), and again prominently at the beginning of any registration or use process (e.g., different models hosted on chatgpt.com). This would, arguably, have a dual beneficial effect. First, it would raise awareness of the energy costs of using models and may incentivise at least a non-negligible fraction of users to refrain from using generative AI models for tasks that can be achieved with similar efficiency with traditional software or search. Second, it would facilitate the comparison between different models that are, for consumers, effectively interchangeable for standardised tasks. Again, an increasing number of consumers may use energy disclosures as a driving element in the choice between different models. In this way, energy information can provide meaningful support for consumers both when deciding if to use generative AI at all and, if yes, which model.

In the EU, such disclosures can be mandated by an update of the Consumer Rights Directive (CRD), for example. The Empowering Consumers for the Green Transition Directive (Directive (EU) 2024/825), adopted in February 2024, specifically amends the CRD to require better information on product durability and reparability, and even compels the display of a 'reparability score' for consumer products (Art. 6(1)(u) and (v) CRD). The extension to an energy score, such as the 'energy efficiency of digital services' (like AI), is a logical next step. A new sub-point to Art. 6(1) should mandate the display of the AI Energy Score 'in a clear and prominent manner' before the consumer is bound by contract.

Third, cloud service providers offering access to AI models and systems should equally be compelled to provide such energy score information, even if they primarily serve business consumers who do not fall under consumer law (such as the CRD). Such a mandate could be included in the revision of the Data Act in the context of the current Digital Omnibus process in the EU, or in the upcoming review of the Digital Services Act. Fourth and finally, any platforms hosting open-source AI models for download or inference should be compelled to prominently display not only the AI Energy Score but also standardised monetary information, to the extent available, such as cost per token or per CO2-equivalent. As the mentioned studies suggest, this not only reinforces the informative power of the energy scores, but also leads customers to make more rational decisions that close the energy efficiency gap.

#### 4.1.1.6 Trade Secret Concerns

Legitimate business interests in protecting proprietary information do not outweigh the public need for environmental transparency in the ways just outlined. Critics might argue that disclosure of model inference consumption and compute location risks trade secret leakage because it reveals hardware and software configurations. However, this type of information proves highly unlikely to make hardware and software choices reverse-engineerable. Multiple external factors beyond those specifications determine inference consumption, e.g.: the model architecture details and parameter counts that remain undisclosed; algorithmic factors, such as the specific optimization techniques employed (e.g., quantization levels, attention compression, or pruning strategies [73]), distillation methods [137], the batch sizes and scheduling algorithms used for query processing [3, 102], the efficiency of the serving infrastructure, as well as the caching strategies and memory management approaches [118]; and also infrastructural components, such as cooling system efficiency and data centre infrastructure design [1]. These variables interact in complex, non-linear ways that make it virtually impossible to deduce specific configurations from aggregate consumption data [cf. 102]. Therefore, trade secret protections do not apply to the disclosures mentioned above.

### 4.1.2 Comparable Metrics and Standardised Benchmarks

Standardised measurement protocols are essential to enable meaningful comparisons between models and track progress toward sustainability goals [16, 61, 65, 97, 110]. Without comparable metrics and benchmarks, model providers can cherry-pick favourable measurements or use incompatible methodologies that obscure true environmental impacts and thwart efforts to compare different models—as is currently the case. Standardization enables market competition based on environmental efficiency, allows regulators to set meaningful benchmarks, and provides researchers with consistent data for longitudinal studies. Reasoning models, which have proliferated in 2025, consume orders of magnitude more energy than standard models—yet this impact correlates poorly with model size, which undermines the common assumption that smaller models are inherently more efficient. Only standardised benchmarking can surface such patterns. Regulators should therefore require providers to submit models for independent benchmarking under a recognised framework, with results made publicly available on centralised leaderboards. More specifically, concerning the two key transparency dimensions suggested (inference and compute location), the following is necessary.

#### 4.1.2.1 For Inference

Inference metrics must capture both standardised performance and real-world usage patterns. Energy usage and water consumption should be measured for certain standardised prompts and benchmarks to enable direct model-to-model comparisons. Again, the AI Energy Score project demonstrates how such standardization can function in practice. It employs custom datasets and identical GPU configurations to benchmark models, which yields results that permit meaningful cross-model comparison. As noted, reasoning-enabled models consume between 150 and 700 times more energy than their non-reasoning counterparts—a finding that would remain invisible without consistent measurement protocols. Regulatory frameworks should require providers to adopt or align with such established benchmarking methodologies rather than permitting each provider to develop proprietary measurement approaches that preclude comparison. To ensure comparability, regulators should adopt open-source standards such as the AI Energy Benchmarks package, which utilises Code Carbon and standardised datasets to measure energy across different hardware configurations.

#### 4.1.2.2 For Compute Locations

Geographic metrics must reveal both typical patterns and distributions of computational resources. For each region, providers should disclose where the majority of compute happens to identify primary processing locations. They should also report how compute distributes on average across facilities within each region to understand redundancy and load balancing, and to enable informed choices about the processing of sensitive data. While transfers of personal data to locations outside of the EU need to be disclosed under the GDPR, such information need not be provided if the data is directly sent, from the consumer device, to a destination outside of the EU (e.g., the US). This direct transmission to a server within the same company does not constitute a transfer in the sense of the GDPR, for lack of a separate importer and exporter of data [33, Example 1]. Moreover, in jurisdictions outside of the EU, such rules may be lacking altogether. Finally, providers should specify what energy mix powers this compute distribution on average to assess carbon intensity.

### 4.1.3 Enforcement

Effective transparency requires multiple enforcement pathways to overcome structural information and power imbalances between model providers and information seekers [13]. This ranges from public to private and collective enforcement [7, 106]. Concerning public enforcement, specialised digital regulators possess the expertise and authority necessary to ensure compliance with transparency requirements. Digital regulators, such as the EU Commission's AI Office for the EU, should oversee compliance through regular audits and investigations. In addition, private enforcement can help. Independent researchers need direct access rights to verify claims and conduct original environmental impact research. Vetted researchers should receive the right to access information about models and compute to derive these data points independently. To this effect, the AI Act should introduce a right comparable to Art. 40 DSA. While the Code of Practice for General-Purpose AI Model Providers foresees certain access rights, their enforcement remains unclear. A normative anchor in the AI Act itself would be preferable. Finally, collective enforcement is starting to become a key anchor of digital regulation [12]. Civil society organizations require standing to represent public interests in environmental transparency. The law should include provisions that allow collective rights organizations, such as consumer organizations, to query this information from GPAI model providers and pursue remedies for non-compliance, as in Art. 80 GDPR.

## 4.2 Right to Use Green Digital Infrastructure

In addition to transparency, users require meaningful control over their engagement with generative AI systems to manage both environmental impacts and concerns about output reliability. This implies making available certain rights to choose environmentally friendly models that are only imperfectly implemented in market settings at the moment. Those rights would address exposure to unwanted or externally imposed generative AI overviews and tools, as well as optimise the choice of relatively environmentally friendly options within the voluntary use of generative AI.

### 4.2.1 Right to Use Search Without Generative AI Overviews

First, search engines must provide users with accessible options to disable AI-generated content and return to traditional search results. Users should be able to easily and persistently implement this choice through a button that turns off AI overviews and other generative AI tools in web search. At present, users can exercise such choice by adding '-ai' to a Google search query, for example. However, many users lack awareness of this option, and it only works for a single query at a time. The interface should make this option prominent and persistent across sessions.

### 4.2.2 Right to Use Digital Infrastructure Without Unnecessary GenAI Components

Second, the principle of user choice over AI integration must extend beyond search to encompass all digital infrastructure services. Increasingly, tools are integrated into software, often by ways of deceptive or dark patterns [46], which nudge users toward harnessing generative AI, e.g. via 'magic wand' AI icons [9]. To counter this tendency, we suggest an encompassing right to use digital infrastructure without unnecessary generative AI. A related provision already exists in the EU. The DSA already establishes a right to non-personalised news feeds against very large online platforms (Art. 38), as recent court proceedings against Meta demonstrate [11]. However, this choice again requires persistence—users currently must select this option every time they open the app according to complaints filed by civil society organizations [34]. A similar pattern is found in the Google Play Store and other apps [9]. Externally imposed AI features reflect a broader need for user sovereignty over AI integration across digital services. Thus, users should be able to persistently use digital infrastructure without unnecessary generative AI components. In this context, unnecessary means that users are typically technically and cognitively able to complete the required task without the use of generative AI, for example in writing text or code.

There are three key reasons for supporting such a user right. First, opting out of generative AI will typically—within certain limits—decrease the carbon footprint of a task such as writing or coding [16]. To this extent, fair market design is coupled with environmental advantages [9]. Second, this right tackles the increasing spread of tools that seek to nudge users to offload mental workload to generative AI instead of thinking, researching or writing for themselves. Such offloading impacts critical thinking [42], a capacity that seems key for genuine progress and democratic discourse. Third, our proposal also draws support from economic theory, specifically the 'penalty default' concept developed by Ayres and Gertner [6]. Their theory suggests that the party that possesses more information should divulge that information to the non-informed party, so that the latter can make an informed decision about what serves their best interest. Hence, the default setting should explicitly disfavour the better-informed party (i.e., penalty default); this mechanism incentivises that party to divulge information to convince the counterparty to alter the arrangement. For instance, Google should explain to users what benefits AI overviews provide; users could then become informed to the extent they desire and opt into AI overview usage. Similarly, users should be able to use WhatsApp without Meta AI integration, and Gsuite tools without email suggestions and other Gemini functionalities. Another example involves software suites that increasingly integrate AI assistants—users should generally retain the ability to use these tools in their traditional form without AI augmentation. One significant challenge emerges: distinguishing technically or functionally necessary from unnecessary use of generative AI may prove difficult, particularly as hybrid or ensemble model architectures increasingly integrate generative AI features into core functionalities. Over time, this could render this right ever more difficult to implement.

### 4.2.3 Right to Use A Green Model

A complementary approach to complete opt-out involves empowering users to choose environmentally optimised AI models when they do voluntarily engage with these systems. Already now, providers typically can deploy a variety of different models to generate text, images, or videos within one model family for each prompt. Many providers already allow users to choose the model they would like to use, though this choice often prioritises performance over environmental considerations. The new right would give users an option to always use a 'green model'—a model that providers optimise for environmental performance. Users should be able to make this choice initially upon registering with the provider and update their model preference at any time. To designate green models, this right could combine with the AI Energy Score framework. A green model would be defined as the model within the model family available on the application that achieves the best energy score rating while still providing the required modality (text, image, video generation, etc.). The Score's star-based system provides a ready-made classification that regulators could incorporate by reference. The AI Energy Score's recent findings underscore why such a right matters: newer models do not automatically prove more efficient, with some 2025 models consuming four times the energy of comparable models from the previous cohort. Without standardised ratings, users cannot identify genuinely efficient options. Moreover, our analysis of reasoning models—which consume up to 700 times more energy than base models—demonstrates that users need clear information to understand when resource-intensive reasoning modes are necessary and when simpler models suffice. Users must have the right to explicitly consent to this 'high-intensity' compute. Platforms should default to 'reasoning off' and require an active user toggle to engage 'reasoning' features, accompanied by a visual indicator of the increased energy cost. This right, combined with non-reasoning default settings, empowers users to align their technology use with their environmental values while maintaining access to AI capabilities. It transforms environmental sustainability from an abstract concern into a concrete choice that users make with each interaction. Moreover, it creates market incentives for providers to develop and deploy more efficient models, as user preferences for green options would drive demand for environmental optimization alongside traditional performance metrics.

## 4.3 International Environmental AI Governance

Finally, the nature of AI's environmental impact demands coordinated international responses that transcend national boundaries and regulatory fragmentation: a global problem requires global solutions. While current geopolitical tensions complicate multilateral cooperation, we must conceptualise robust international frameworks for future implementation when political conditions prove more favourable. The challenge remains worth addressing despite present obstacles because AI's environmental effects ignore national borders—emissions from data centres in one country affect global climate patterns, and water consumption in drought-prone regions has transnational implications.

Ideally, a Global AI Compact should incorporate the rights and disclosures contemplated above, creating uniform standards for environmental transparency and user empowerment across jurisdictions. This compact would prevent regulatory arbitrage where providers relocate operations to jurisdictions with weaker environmental standards, and it would establish baseline protections that users enjoy regardless of their location. The UN would provide an appropriate forum to coordinate and implement such a framework, with specialised agencies such as the International Telecommunication Union (ITU) offering technical expertise and existing governance structures. Other UN bodies, such as the United Nations Environment Programme (UNEP) and the Framework Convention on Climate Change (UNFCCC) secretariat, could contribute environmental expertise and integrate AI considerations into broader climate governance mechanisms.

At a minimum, even outside of traditional international law organizations, a bottom-up framework on environmental impact measurement should be developed by scholars and practitioners. This framework would enable consistent cross-border assessment of AI's environmental footprint, facilitate comparative research across different regulatory environments, and provide a technical foundation that future international agreements could adopt. Academic institutions, industry consortia, and civil society organizations could collaborate to establish measurement protocols, share best practices, and build consensus around environmental standards even absent formal treaty obligations. Such grassroots efforts often precede and inform eventual international legal frameworks, as witnessed in domains from internet governance to sustainable development standards.

Finally, a 'coalition of the willing,' consisting of countries and regions with regulatory power, could follow the Irish example (see 3.1) and adopt administrative decisions which mandatorily link the construction or grid connection of data centres to sustainability targets, such as additional renewable energy quota. In this way, individual states can reassert control over and ownership of their technology future on a local level with lasting global impact.

## 5 Conclusion

Our main contributions are threefold. First, we document that new modalities of AI models, such as reasoning models and generative AI employed in Web search, multiply energy consumption by significant factors, ranging from 150-700 for reasoning models.

Second, our comparative legal analysis suggests that the overall increase in environmental strains that AI is putting on the planet is, paradoxically, accompanied by a decrease in legally mandated transparency at the global level. Only the EU has enacted specific, model-level disclosure obligations. However, they remain limited as they do not squarely address inferences, and are not disclosed to NGOs, researchers, or the public at large.

Third, we therefore make several policy proposals for meaningful transparency, user empowerment, and international coordination. An important desideratum is to shift policy discourse and concrete provisions away from mere facility-level disclosure obligations, focused on data centres, to additional model-level transparency. This is crucial for market participants, researchers, and the public at large to become informed about the increasing environmental costs of AI as generative models increasingly are pushed into not only web search and checkbox, but almost any imaginable software interface. As we detail, such disclosures must include not only training but also inference components, and compute locations, and be paired with strategic visual labels to guide user and consumer choice. However, transparency alone is not enough. Often, environmentally-aware consumers lack meaningful choice. Hence, we advocate for two key rights concerning green digital infrastructure: a right to use digital infrastructure without unnecessary GenAI components to counter the ubiquitous and non-consensual insertion of generative AI tools; and a right to a green model for cases of voluntary AI use. Finally, we chart top-down and bottom-up pathways for a strengthened international framework of environmental AI governance. Following the Irish example, we also recommend the use of administrative decision-making, for example concerning grid connections, for promoting sustainability targets where fully legislative proposals may fail. Despite the politically difficult times, global solutions remain more necessary than ever.

## A. Appendix: Additional Comparative Analysis

Table 1. Regulatory transparency and climate policy comparison

| | EU | USA | CHN | GBR | JPN | CAN | IND | BRA | UAE | SGP |
|---|---|---|---|---|---|---|---|---|---|---|
| GHG emissions reporting for specific industries (varying scope and threshold) | **yes** | not en-forced | **yes** | **yes** | **yes** | **yes** | **yes** | **yes** | **yes** | **yes** |
| ICT-sector-specific mandatory transparency | **yes** | **no** | **no** | **no** | **yes** | **no** | **no** | **no,** pro-posed | **no** | **no** |
| Model-level transparency | **yes** | **no** | **no** | **no** | **no** | **no** | **no** | **no** | **no** | **no** |
| Share of renewable electricity generation in 2024 [56] | 48% | 23% | 34% | 52% | 23% | 65% | 22% | 87% | 11% | 5% |
| Carbon zero commitments [15] | 2050 | no target | 2060 | 2050 | 2050 | 2050 | 2070 | 2050 | 2050 | 2050 |

### A.1 The United Kingdom

The UK's approach to energy reporting and transparency for data centres and AI remains largely hands-off and focused on enabling infrastructure growth rather than imposing sector-specific disclosure requirements. Key policy documents—the 2021 National AI Strategy [22], the 2023 AI Regulation White Paper [24], and the 2025 AI Opportunities Action Plan [25]—prioritise innovation, compute expansion, and related energy-capacity, with limited action regarding other energy and emissions impacts of AI, even after data centres were designated critical national infrastructure in 2024 [50]. In practice, data centres are covered mainly by general building efficiency rules, such as green building standards and Minimum Energy Efficiency Standards, which do not adequately reflect their energy intensity [101]. Corporate-level transparency is provided through the Streamlined Energy and Carbon Reporting framework, requiring large companies to disclose energy use and emissions, but without AI- or data-centre-specific metrics [23]. Voluntary schemes, notably Climate Change Agreements that offer discounts on the Climate Change Levy in exchange for efficiency commitments, provide additional but limited incentives, with a new scheme under consideration from 2027 [49]. Beyond these measures, the government has largely confined its role to researching the impacts of data-centre growth and considering potential policy responses to energy challenges [50]. The regulatory logic is, hence, one of facility-level regulation with additional voluntary guidance and overall lower environmental ambition.

### A.2 India

The 2001 Energy Conservation Act allows the Central government, in consultation with the Bureau of Energy Efficiency (BEE), to classify certain entities as energy-intensive and prescribe efficiency standards [87]. This includes Renewable Consumption Obligation (RCO) rules which were updated in September 2025 and allow to set a certain renewable energy factor, however, data centre operators have not yet been designated. On the State level, several data centre policies have been published that aim to facilitate renewable-energy procurement but do not mandate efficiency standards [68]. Overall, there has not yet been a material consensus on a unified policy response to the environmental effects of AI and data centre operations in India.

### A.3 Brazil

Brazil is promoting itself as a data-centre and AI hub by leveraging its electricity grid, which is about 87 % renewable, while gradually introducing energy-efficiency and transparency requirements. The National AI Plan (May 2024) allocates roughly R$23 billion to AI infrastructure, including a R$2 billion credit line for data centres [81], and recent tax incentives under Measure 1318/2025 (ReData) are explicitly conditional on environmental commitments such as 100 % renewable electricity use, water-use-efficiency metrics and enhanced reporting from 2026 [82]. Although comprehensive regulation is still pending, proposed legislation—notably Bill No. 3018/2024—would require data-centre operators to report annual energy consumption, conduct energy audits, use renewables and set emissions-reduction targets, with tax benefits tied to environmental certification [72]. The overall regulatory strategy is, hence, one of voluntary disclosures and plans for facility-level regulation with a higher environmental ambition.

### A.4 The United Arab Emirates

Following its declaration of 2023 as the ‘year of sustainability’ [86], the United Arab Emirates (UAE) have increased their focus on sustainability in the digital and AI sectors. The 2024 Federal Decree-Law No. 11 [119] introduced mandatory emissions reporting for entities with Scope 1 and 2 emissions exceeding 0.5 $MtCO2e$, which will also include large data centre operators. Sector-specific direction comes from policy guidance, notably the 2024 General Framework for Adopting Sustainable Digital Transformation [48], which prioritises green data centres and cloud computing, and the Ministry of Finance’s 2025 Sustainable Digital Services (IT) Guideline [85], which encourages voluntary energy and emissions transparency through compliance with international standards such as ISO 50001, ASHRAE 90.4, and ISO 14064-1 and the GHG Protocol. Growing electricity demand from large AI data centres, such as OpenAI’s planned 1 GW Stargate UAE cluster, with a potential expansion to 5 GW, has also prompted the Ministry of Energy and Infrastructure to establish a national review team [43]. The regulatory approach is, therefore, one of limited facility-level transparency, mostly voluntary guidance, and medium environmental ambition.

### A.5 Canada

Canada currently has no AI- or data centre-specific mandatory transparency requirements. Federal policy instead focuses on system-wide decarbonisation through the Clean Electricity Regulations (in force January 2025) [32] and the clean electricity strategy Powering Canada’s

Future (October 2024) [93], which commit to a net-zero grid by 2050 and acknowledge AI as both a grid optimisation tool and a growing source of electricity demand, while the proposed Artificial Intelligence and Data Act (AIDA) has stalled [5]. In response to surging data-centre demand that exceeds grid capacity, provinces are asserting control through connection prioritisation and restrictions—such as Ontario's 2025 Electricity Act amendments [84] and British Columbia's 2025 limits on crypto-mining and AI data centres [111]—rather than through transparency mandates. Energy efficiency and disclosure expectations are therefore largely voluntary, guided by NRCan's Best Practice Guide for Canadian Data Centres (November 2024) [92]. The overall policy response is focused on resource distribution challenges while existing guidance on transparency is voluntary and limited to the facility-level. The overall environmental ambition, therefore, seems medium taking into account the overall net-zero target but a lack of sector-specific hard regulation.

### A.6 Singapore

Singapore's energy reporting and transparency requirements for data centres, including AI-driven facilities, are largely governed by economy-wide sustainability regulations rather than AI-specific rules. Under the Energy Conservation Act [91], large energy-intensive entities must report energy use and implement efficiency measures, while facilities emitting at least 25,000 $tCO_2e$, are subject to the carbon tax [89]. These obligations are complemented by sector-specific roadmaps and voluntary standards—such as the 2024 IMDA Green Data Centre Roadmap [54], the Green Mark certification for data centres [52], the Tropical Data Centre Standard [53], and the Energy-Efficiency Standard for Data-Centre IT-Equipment [55]—which promote greater transparency on data centre energy performance through efficiency benchmarks, equipment standards, and operational guidance. The regulatory logic, hence, is one of limited facility-level regulation, paired with extensive voluntary guidance and an overall medium environmental ambition.

### A.7 International Cooperation

The UN has been active on the issue of energy consumption of AI through reports, recommendations, and summits—such as UNESCO's AI Ethics Recommendations [120], the ITU report on AI and the environment [60], and initiatives under the Global Digital Compact and UNFCCC—but these efforts have not led to concrete cooperation or binding obligations, reflecting persistent geopolitical constraints.

The OECD and GPAI, after earlier work on AI principles and environmental impact measurement, have become less active in this area in recent years [94, 96, 98]. By contrast, ISO has more recently taken an operational role by developing standards and reports relevant to energy and sustainability transparency, including ISO/IEC 42001 on AI management systems, ISO/IEC 42005 on impact assessment, and ISO/IEC TR 20226:2025, which sets out methodologies for measuring and reporting the environmental impacts of AI and data centres [57–59]. However without any binding international cooperation, companies have few incentives to follow through on energy transparency.